\pdfoutput=1

\documentclass[a4paper]{article}

\usepackage{INTERSPEECH2020}
\usepackage{amsmath}
\usepackage{CJKutf8}
\usepackage{bbding}
\usepackage{color}
\usepackage{multirow}
\usepackage{arydshln}
\usepackage{booktabs}

\title{Semantic Data Augmentation for End-to-End \\ Mandarin Speech Recognition}
\name{Jianwei Sun, Zhiyuan Tang, Hengxin Yin, Wei Wang, Xi Zhao, Shuaijiang Zhao, Xiaoning Lei, Wei Zou, Xiangang Li}
\address{
   KE Holdings Inc, Beijing, China}
\email{\{sunjianwei006, tangzhiyuan001, zhaoshuaijiang001, zouwei026, lixiangang002\}@ke.com, {hengxinyin@csu.edu.cn, 2181002081@cnu.edu.cn, zxcdlucy@163.com}}

\ninept

\begin{document}
\begin{CJK*}{UTF8}{gbsn}

\maketitle

\begin{abstract}
End-to-end models have gradually become the preferred option for
automatic speech recognition (ASR) applications.
During the training of end-to-end ASR,
data augmentation is a quite effective technique for regularizing
the neural networks.
This paper proposes a novel data augmentation technique based on
semantic transposition of the transcriptions via syntax rules for end-to-end Mandarin ASR.
Specifically, we first segment the transcriptions based on part-of-speech tags.
Then transposition strategies, such as placing the object in front of the subject or swapping the subject and the object, are applied on the segmented sentences.
Finally, the acoustic features corresponding to the transposed transcription are
reassembled based on the audio-to-text forced-alignment produced by a pre-trained ASR system.
The combination of original data and augmented one is used for training a new ASR system.
The experiments are conducted on the Transformer\cite{dong2018speech}
and Conformer\cite{gulati2020conformer} based ASR.
The results show that the proposed method can give consistent performance gain to the system.
Augmentation related issues, such as comparison of different strategies and ratios for data combination are also investigated.

\end{abstract}

\section{Introduction}

In recent years, end-to-end models have become a primary paradigm for
automatic speech recognition (ASR) applications,
generally under the framework of Connectionist Temporal Classification (CTC) loss function\cite{graves2014towards,hannun2014deep},
attention mechanism\cite{chorowski2015attention,chan2015listen,bahdanau2016end}
or joint CTC-attention\cite{kim2017joint} alongside with a variety of neural network structures,
e.g., LSTM\cite{graves2013hybrid}, Transformer\cite{dong2018speech} and Conformer\cite{gulati2020conformer}.
The training of end-to-end ASR systems usually consumes a large amount of data
to avoid overfitting, while it's unreasonable to obtain too much
expensive human-labeled speech data.
Data augmentation is a quite promising approach to alleviate this problem,
by increasing the quantity of speech data simply based on existed materials.


Many data augmentation methods have been proposed for ASR,
typically conducted on either acoustic or linguistic level.
Acoustic augmentation usually increases the variations of existed
acoustic signals or features via strategies such as
vocal tract length perturbation\cite{jaitly2013vocal},
noise insertion\cite{seltzer2013investigation},
speed perturbation\cite{ko2015audio},
acoustic room simulator\cite{kim2017generation}
and SpecAugment\cite{park2019specaugment}.
The resulted acoustic inputs share the same labels as original.
On the other hand, linguistic augmentation generates new transcriptions firstly,
then obtains corresponding acoustic inputs by splicing or synthesis.
For example, neural text-to-speech synthesis
has been widely used as a data augmentation approach for ASR,
by introducing more acoustic and linguistic diversity necessary for
robustness or leveraging more training data for low-resource scenario
\cite{rosenberg2019speech,gokay2019improving,rossenbach2020generating,du2020speaker}.
As in the work of \cite{du2021data},
code-switching scenario was investigated and audios were spliced based on phone alignment,
and also synthesized with new text generated by word translation or word
insertion.
When it comes to Mandarin ASR, the linguistic augmentation methods
can be very different because of the specific language properties owned by Chinese.
For example, Chinese verb never changes its form regardless of any tenses and
there is no distinction between singular and plural for
Chinese noun or pronoun.
Those specific properties give many inspirations in the developing of
linguistic-level data augmentation for Mandarin ASR.

In this paper, we propose a simple yet effective data augmentation technique based on
semantic transposition of the transcriptions via syntax rules for end-to-end Mandarin ASR.
Precisely, the transposition strategies are designed based on Chinese syntax rules
to make sure the resulted sentences still conform to formal syntax, regardless of
whether the original meaning changes or not.
Some transposition strategies are presented, such as
placing the object in front of the subject or swapping the subject and the object.
For the process of data augmentation,
the transcriptions are firstly segmented based on part-of-speech tags,
then selected transposition strategies are conducted on them.
Finally, the acoustic features corresponding to the transposed transcription are
reassembled based on the frame level forced-alignment produced by a pre-trained GMM-HMM ASR system.
The combination of original data and augmented one is used for training a new end-to-end ASR system.
Although the splicing process of new acoustic inputs is similar to that in the work of \cite{du2021data},
the latter one simply generates new texts by replacing the English segment of
one code-switching utterance with that of another utterance,
and our proposed method produces new texts by exchanging places of words with more flexible rules inside the single utterance.

The rest of the paper is organized as follows:
Section \ref{sec:method} describes the details about semantic data augmentation,
including the designing of transposition strategies and the process of data augmentation.
Section \ref{sec:exp} reports the experimental performance,
and introduces some analysis including the topic of visualization,
comparison of different rules and combination ratios between augmented data and original one.
The paper is concluded in Section \ref{sec:conc} with some future plan.

\section{Semantic Data Augmentation}
\label{sec:method}

\subsection{Chinese Transposition Strategies}

The syntax rules of Mandarin Chinese are flexible, but they share a basic rule which can be summarized 
as ``(attribute) subject + [adverbial] predicate $\langle$complement$\rangle$ + (attribute) object"\cite{shi2002establishment,ross2017modern}.
In daily communication, Mandarin speakers often use some syntax rules to change the order of words in the sentence, such as inverted sentence.
The common syntax rules of Mandarin are showed in Table \ref{tab:tab1}.
For example, “我今天要去公园” (I am going to the park today) can be changed to “公园我今天要去” (I am going to the park today) through the syntax rule R2, and “我今天很高兴” (I am very happy today) can be changed to “我很高兴今天” (I am very happy today) through the syntax rule R7. 
Some rules keep the original meaning of the utterance after transposition, like inverted sentence,
while others may change the meaning, providing more acoustic and linguistic diversity for model training.
For example, “我很喜欢朋友” (I like my friends very much) can be changed to “朋友很喜欢我” (Friends like me very much) via the syntax rule R1.
Furthermore, if the attributes of the subject and the object are exchanged with each other, the semantics of the sentence itself will be enriched. For example, the sentence “我很帅他很丑” (I am handsome and he is ugly) can be change to “我很丑他很帅” (I am ugly and he is handsome).
There are also some strategies that make the meaning of resulted text irrational but still
meeting the syntax rules, e.g., “我要去公园” (I am going to the park) can be changed to “公园要去我” (The park is going to me) by syntax rule R5. 

The Mandarin text can be easily expanded in terms of sentence structure and semantics through the syntax rules, alleviating the lack of training corpus to a certain extent.
In this paper, the part-of-speech tagging tool CoreNLP\cite{manning2014stanford}
is used to segment and tag the text. Based on the presented syntax rules in Table \ref{tab:tab1}, the sentence pattern and semantics of the text are expanded to enrich our linguistic corpus.

\begin{table}[th]
  \caption{Common syntax rules of Mandarin Chinese}
  \label{tab:tab1}
  \centering
  \begin{tabular} {p{15pt}p{180pt}}
    \toprule
    \multicolumn{1}{l}{\textbf{Rules}} & \multicolumn{1}{l}{\textbf{Details}} \\
    \midrule
    R1  & Switch the position of subject and object                \\
    R2  & Put the object at the beginning of the sentence      \\
    R3  & Put the attributive at the beginning of a sentence   \\
    R4  & Switch the positions of adjectives and adverbs next to each other                               \\
    R5  & Put predicate at the beginning of a sentence                   \\
    R6  & Swap the positions of two adjectives \\
    R7  & Put the adverbial at the beginning of a sentence         \\
    \bottomrule
  \end{tabular}
\end{table}

\subsection{Augmentation Procedure}
\begin{figure}[thb]
  \centering
  \includegraphics[width=\linewidth, trim=0 0 0 0]{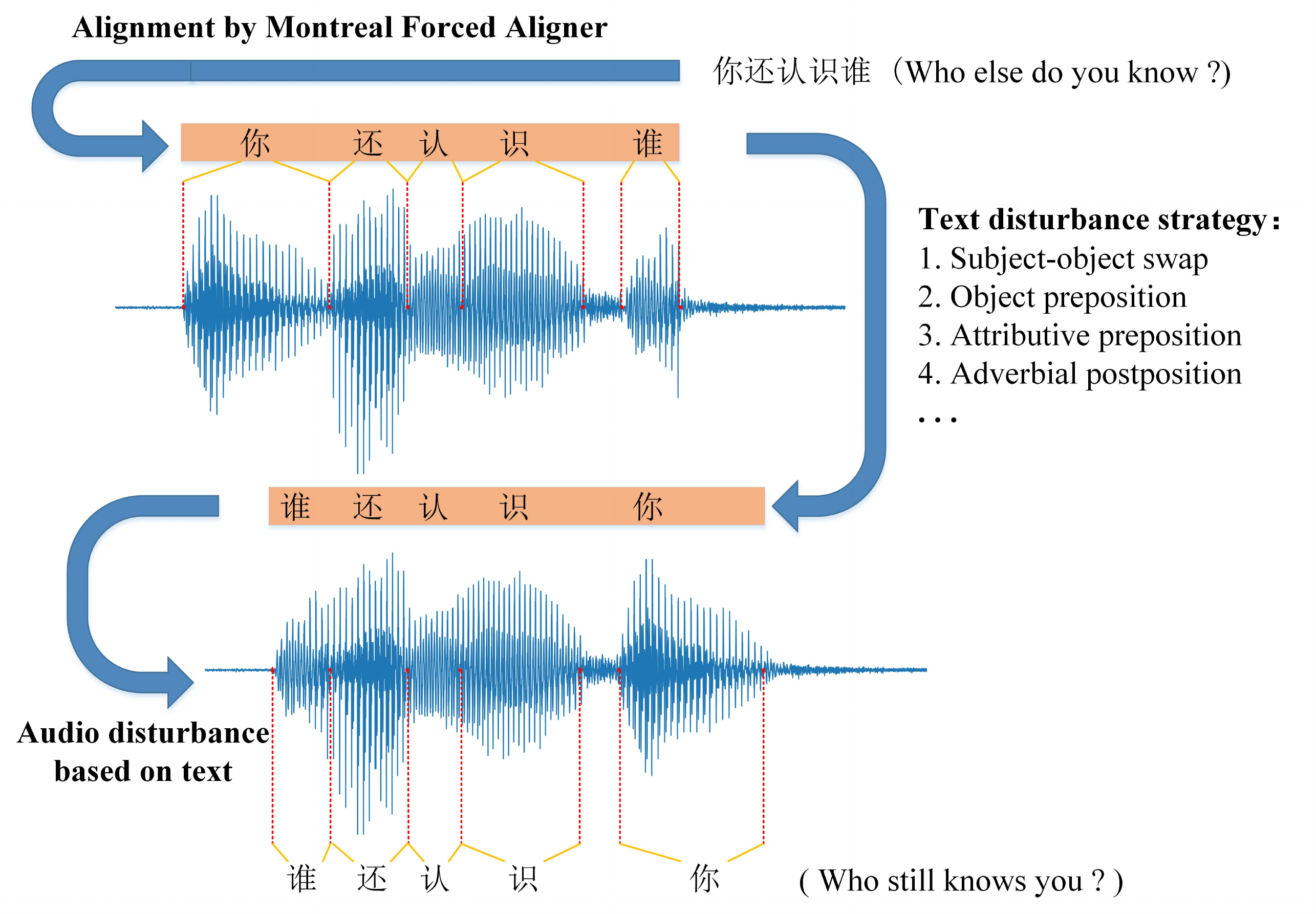}
  \caption{The flow-chart of Semantic Data Augmentation.}
  \label{fig:example_semantic}
\end{figure}

After transposing existed transcriptions,
related speech features are resembled based on frame-level forced-alignment produced by a pre-trained ASR system.
The details involve 3 steps as shown in Figure \ref{fig:example_semantic}.

\subsubsection{Framewise forced-alignment}
In order to get acoustic features corresponded to the transposed transcriptions,
in this paper we resemble the existed features based on framewise forced-alignment.
Precisely, we can extract the acoustic segments associated with any specific words via forced-alignment,
and then resulted acoustic segments in one sentence can be reordered according to the transposition of its text.
There are a variety of methods to align transcriptions and acoustic features.
Moreno et al. \cite{moreno1998recursive} obtained text-to-speech alignment by selecting words as anchor points.
Similarly, Haubold and Kender \cite{haubold2007alignment} used a set of small acoustic units as anchor points to align audio and text.
Hoffmann and Pfister \cite{hoffmann2013text} proposed an HMM-based sentence-level forced-alignment method through a phoneme model set.
Ahmed et al. \cite{ahmed2013technique} used syllable information in speech and text to achieve sentence-level alignment.
Martin \cite{martin2004winpitchpro} used a new generation of acoustic analysis software to complete the alignment of text and audio.
In this paper, we apply commonly used GMM-HMM ASR system to achieve forced-alignment\cite{dahl2011context}.

\subsubsection{Part-of-speech tagging and text expanding}
The part-of-speech tagging tool CoreNLP performs word segmentation on the text.
Sentence components of texts are analyzed based on the part-of-speech tags.
Table \ref{tab:word_styles} shows the mapping relationship between the part-of-speech tags and sentence patterns.
According to Table \ref{tab:word_styles}, the sentence pattern corresponding to each text can be found, and then the text is transposed according to the syntax rules as shown in Table \ref{tab:tab1}. If the part-of-speech tag of the sentence is only a single noun or adjective without complete sentence patterns, the text should not be transposed to avoid semantic confusions or errors.

\subsubsection{Feature resembling}
After completing semantic augmentation for linguistic data,
the segments of audios are reordered and resembled accordingly.
Finally, the augmented data will be combined with existed data to train a new ASR model.

\begin{table*}[t]
  \caption{The mapping relationship between part-of-speech tags and sentence patterns}
  \label{tab:word_styles}
  \centering
  \begin{tabular} {p{150pt}p{250pt}}
    \toprule
    \textbf{Part-of-speech}      &\textbf{Sentence pattern}  \\
    \midrule
    (noun) + (verb) + (noun)                    & subject (noun) + predicate (verb) + object (noun)                   \\
    (noun) + (adverb) + (adverb) + (adjective)  & subject (noun) + adverbial (adverb + adverb) + attribute (adjective) \\
    (noun) + (adverb) + (verb) + (noun)         & subject (noun) + adverbial (adverb) + predicate (verb) + object (noun) \\
    (noun)                                      & single noun                               \\
    (adjective)                                 & single adjective                   \\
    \bottomrule
  \end{tabular}
\end{table*}

\section{Experiment}
\label{sec:exp}

\subsection{Dataset}
We perform the experiments on two corpora: HKUST and AISHELL-1. The HKUST was collected and transcribed by Hong Kong University of Science and Technology (HKUST), which contains 873 calls (about 150 hours) in training set, 4000 utterances in development set and 24 calls (about 5,000 utterances) in test set. The AISHELL-1 contains 120,098 utterances in training set, 14,326 utterances in developments set and 7,176 utterances in test set.

All experiments are conducted using 40-dimensional log-Mel filter bank features, computed with a 25ms window and shifted every 10ms. The features are normalized via global mean subtraction and global variance normalization.

\subsection{Model}
We use end-to-end ASR based on two promising neural structures, i.e. Transformer and Conformer, to verify our proposal
with the tool Espnet\cite{mehta2018espnet}.

For the two self-attention based encoder-decoder models,
the encoder includes 12 layers and the decoder 6 layers.
Besides, the number of filters in convolutional network for Conformer is 512,
the head number of multi-head attention is 4,
the dimension of input and output in feedforward network is both 512,
and the dimension of inner layer in feedforward network is 2,048.
The language models for Transformer and Conformer are RNN-LM and Transformer-LM respectively.

Warming-up learning rate strategy is applied to train both Transformer and Conformer.
Besides, we train the Transformer models 20 epochs in total with batch size 32 and train the Conformer models 50 epochs in total with batch size 32.
Considering tractability and rationality in the text transposition,
the first four syntax rules showed in Table \ref{tab:tab1} are selected to
do semantic data augmentation.




\subsection{{Performance}}

\begin{table*}[th]
  \caption{CER($\%$) with or without semantic data augmentation (SA) using different acoustic and language models.}
  \label{tab:CER2}
  \centering
  \begin{tabular}{p{85pt}p{45pt}p{15pt}p{15pt}p{0.1pt}p{15pt}p{15pt}p{15pt}p{15pt}}
  \toprule
    \multicolumn{1}{l}{\textbf{Acoustic model}}  &\multicolumn{2}{l}{\textbf{Language model}} &&
    \multicolumn{2}{l}{\textbf{HKUST}} & & \multicolumn{2}{l}{\textbf{AISHELL-1}} \\
    \cline{2-3}
    \cline{5-6}
    \cline{8-9}
     & \multicolumn{1}{l}{\textbf{structure}} & \multicolumn{1}{l}{\textbf{text}} &
      &\multicolumn{1}{l}{\textbf{dev}} & \multicolumn{1}{l}{\textbf{test}} &&
    \multicolumn{1}{l}{\textbf{dev}} & \multicolumn{1}{l}{\textbf{test}} \\
    \midrule
    Transformer\cite{transformer2019text} &RNN & Raw & & None & 23.7 && 6.0 & 6.7 \\
    \cdashline{1-9}[1pt/2pt]
    \multirow{3}{*}{Transformer(ours)} & \multirow{3}{*}{RNN}
      & None               && 22.1 & 21.8 && 6.5 & 7.0 \\
     & &Raw	  && 19.8 & 20.9 && \textbf{\textbf{6.1}} & 7.0 \\
     &  &SA     && 21.8 & 20.8 && 6.3 & 6.8 \\
    \cdashline{1-9}[1pt/2pt]
    \multirow{3}{*}{Transformer-SA(ours)} &\multirow{3}{*}{RNN}   & None               && 20.8 & 21.9 && 6.3 & 7.1 \\
    &  & Raw   && 20.4 & 21.1 && 6.3 & 6.9 \\
      &  &SA     && \textbf{\textbf{19.4}}& \textbf{\textbf{20.7}} && \textbf{\textbf{6.1}}& \textbf{\textbf{6.6}}\\
    \midrule
    Conformer(Espnet) &Transformer & Raw & & None & None && 4.3 & 4.6 \\
    \cdashline{1-9}[1pt/2pt]
    \multirow{3}{*}{Conformer(ours)} & \multirow{3}{*}{Transformer}
     & None                 && 16.7 & 18.6 && 5.0 & 5.5 \\
     & &Raw	      && 16.3 & 18.5 && 4.3 & 4.6 \\
     & &SA         && 16.6 & 18.6 && 5.0 & 5.5 \\
     \cdashline{1-9}[1pt/2pt]
      \multirow{3}{*}{Conformer-SA(ours)} &\multirow{3}{*}{Transformer}
      & None                 && 16.6 & 18.6 && 4.5 & 4.9 \\
       & &Raw       &&16.4& 18.5 && 4.5 & 4.9 \\
      & &SA  && \textbf{\textbf{15.9}}& \textbf{\textbf{18.3}} && \textbf{\textbf{4.2}} & \textbf{\textbf{4.5}} \\

    \bottomrule
  \end{tabular}
\end{table*}

Table \ref{tab:CER2} shows the Character Error Rate (CER) results on the test sets of HKUST and AISHELL-1 with both Transformer and Conformer, including the state-of-the-art results from Espnet.
In addition to different neural structures for acoustic and language models,
we also investigate the influence of semantic data augmentation on both components respectively,
i.e., both components can be trained with original data or augmented data.
For end-to-end ASR, language model can even be omitted.

It can be seen that the best results are obtained with
our proposed semantic data augmentation (noted as SA),
and the performance can be even improved when the augmentation is applied on either the acoustic model or the language model.
It can be explained that semantic data augmentation expands both acoustic and linguistic data, which gives diversity on different levels and the improvements stack on each other.



\subsection{Analysis}
\label{sec:ana}

\subsubsection{Visualization}
\begin{figure}[thb]
  \centering
  \includegraphics[width=\linewidth, trim= 0 0 0 0]{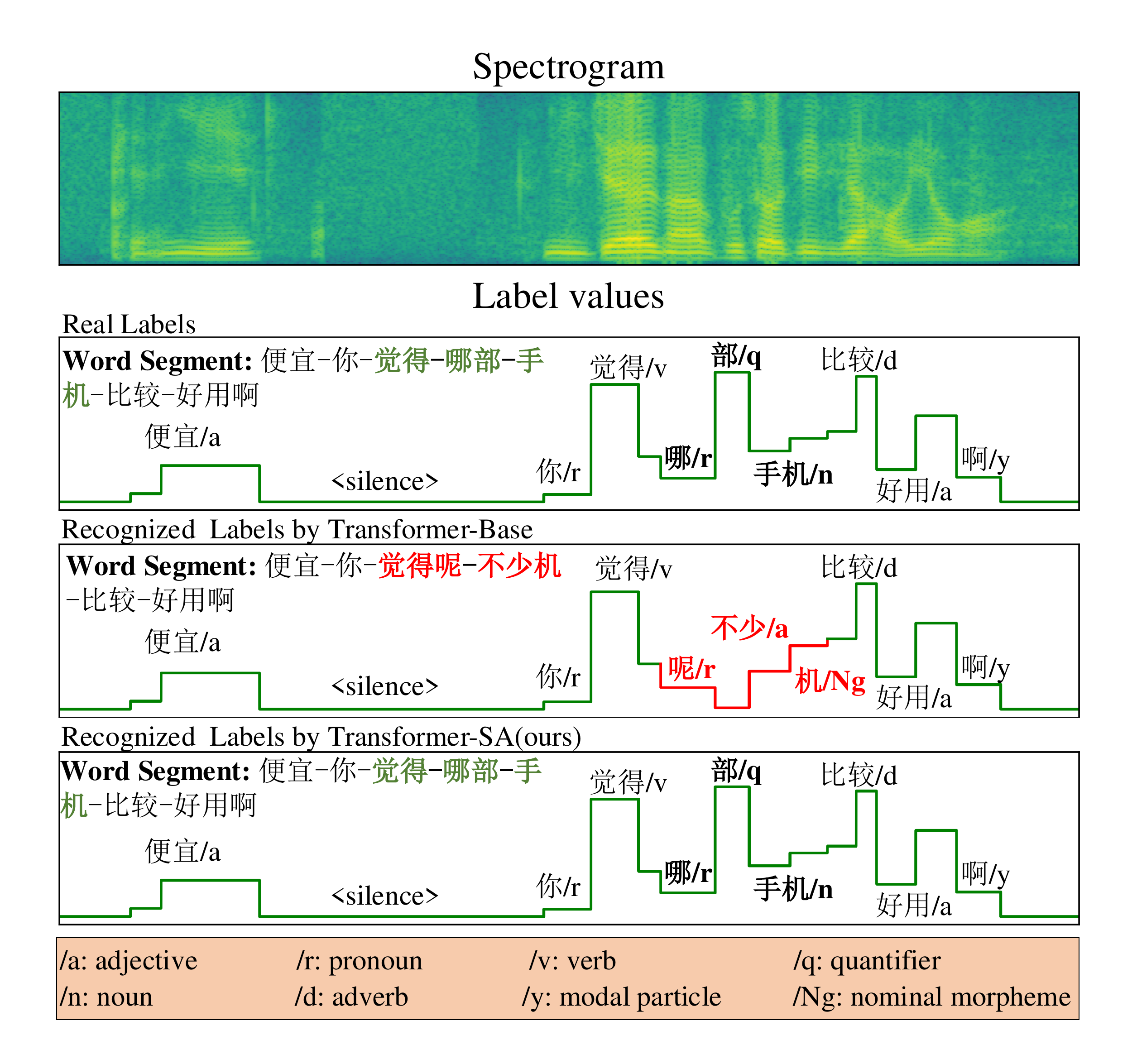}
  \caption{Typical test sample with recognized labels by Transformer-based ASR with or without semantic data argumentation (noted as SA).}
  \label{fig:fig2}
\end{figure}

To better understand how the proposed augmentation improves the system's performance,
we visualize the spectrogram of a test sample and its labels alongside with recognized ones by Transformer-based ASR
with or without semantic data augmentation, as shown in Figure \ref{fig:fig2}.
We segment the ground-truth labels and recognized ones through the part-of-speech tagging tool CoreNLP for better comparison.
We can see that the baseline model recognizes “觉得-哪部-手机” as “觉得呢-不少机”, and does not accurately recognize the relatively complex part-of-speech of the sentence, while the model with semantic data augmentation can better identify the phrases such as “哪部” and “手机”.
Also, spectrogram shows more segments along the horizontal time axis, indicating more complex syntax structure,
so more similar data for training is better for recognizing this kind of samples.
Overall, the improvement can be attributed that
the proposed augmentation method provides more diversity on both acoustic and linguistic level.

\subsubsection{Comparison of rules}
We compare the impact of the selected four syntax rules for semantic data augmentation algorithm on performance of both Transformer and Conformer, as shown in Table \ref{tab:tab5} and Table \ref{tab:tab6}, respectively. It can be seen that different syntax rules have different effects on the performance of different models.
None of the four rules can improve the system consistently, even for the same model,
while when applying the four syntax rules together,
consistent performance gains are obtained.
We argue that the applying of only a single rule may destroy the balance of different syntax patterns,
but more syntax rules can give more syntax diversity.

\begin{table}[th]
  \caption{CER($\%$) of four different syntax rules for semantic augmentation algorithm on Transformer.}
  \label{tab:tab5}
  \centering
  \begin{tabular}{p{12pt}p{12pt}p{12pt}p{12pt}p{18pt}p{18pt}p{0.1pt}p{18pt}p{18pt}}
  \toprule
    \multicolumn{1}{c}{\textbf{R1}} & \multicolumn{1}{c}{\textbf{R2}} &
    \multicolumn{1}{c}{\textbf{R3}} & \multicolumn{1}{c}{\textbf{R4}} &
    \multicolumn{2}{l}{\textbf{HKUST}} && \multicolumn{2}{l}{\textbf{AISHELL-1}} \\
    \cline{5-6}
    \cline{8-9}
    & & & & dev & test && dev & test \\
    \midrule
    \XSolid & \XSolid & \XSolid & \XSolid             & 19.8 & 20.9 &
                                                      & \textbf{\textbf{6.1}} & 7.0 \\
    \Checkmark & \XSolid & \XSolid & \XSolid          & 20.1
                                                      & \textbf{\textbf{20.7}} &
                                                      & 6.2
                                                      & \textbf{\textbf{6.5}} \\
    \XSolid & \Checkmark & \XSolid & \XSolid          & 20.2 & 20.9 && 6.5 & 6.7 \\
    \XSolid & \XSolid & \Checkmark & \XSolid          & 20.1 & 21.0 && \textbf{\textbf{6.1}} & 7.0 \\
    \XSolid & \XSolid & \XSolid & \Checkmark          & 19.8 & 20.9 && \textbf{\textbf{6.1}} & 6.6 \\
    \Checkmark & \Checkmark & \Checkmark & \Checkmark & \textbf{\textbf{19.4}}
                                                      & \textbf{\textbf{20.7}} &
                                                      & \textbf{\textbf{6.1}} & 6.6 \\
    \bottomrule
  \end{tabular}
\end{table}

\begin{table}[th]
  \caption{CER($\%$) of four different syntax rules for semantic augmentation algorithm on Conformer.}
  \label{tab:tab6}
  \centering
  \begin{tabular}{p{12pt}p{12pt}p{12pt}p{12pt}p{18pt}p{18pt}p{0.1pt}p{18pt}p{18pt}}
  \toprule
    \multicolumn{1}{c}{\textbf{R1}} & \multicolumn{1}{c}{\textbf{R2}} &
    \multicolumn{1}{c}{\textbf{R3}} & \multicolumn{1}{c}{\textbf{R4}} &
    \multicolumn{2}{l}{\textbf{HKUST}} && \multicolumn{2}{l}{\textbf{AISHELL-1}} \\
    \cline{5-6}
    \cline{8-9}
    & & & & dev & test && dev & test \\
    \midrule
    \XSolid & \XSolid & \XSolid & \XSolid             & 16.3 & 18.5 && 4.3 & 4.6 \\
    \Checkmark & \XSolid & \XSolid & \XSolid          & 15.7 & 18.5 &
                                                      & \textbf{\textbf{4.2}} & 4.6 \\
    \XSolid & \Checkmark & \XSolid & \XSolid          & \textbf{\textbf{14.7}} & 18.7 &
                                                      & \textbf{\textbf{4.2}} & 4.6 \\
    \XSolid & \XSolid & \Checkmark & \XSolid          & 16.3
                                                      & 18.6 && 4.3 & 4.6 \\
    \XSolid & \XSolid & \XSolid & \Checkmark          & 16.6 & 19.2 && 4.3 & 4.7 \\
    \Checkmark & \Checkmark & \Checkmark & \Checkmark & \textbf{\textbf{15.9}}
                                                      & \textbf{\textbf{18.3}} &
                                                      & \textbf{\textbf{4.2}}
                                                      & \textbf{\textbf{4.5}} \\
    \bottomrule
  \end{tabular}
\end{table}

\subsubsection{Data combination ratio}

\begin{table}[th]
  \caption{Comparison of CER(\%) under different data combination ratios with Transformer.}
  \label{tab:tab7}
  \centering
  \begin{tabular}{p{8pt}p{8pt}p{8pt}p{8pt}p{8pt}p{0.1pt}p{8pt}p{8pt}p{0.1pt}p{8pt}p{8pt}}
  \toprule
    \multicolumn{5}{l}{\textbf{Data combination ratios}} &&
    \multicolumn{2}{l}{\textbf{HKUST}} && \multicolumn{2}{l}{\textbf{AISHELL-1}} \\
    \cline{1-5}
    \cline{7-8}
    \cline{10-11}
   Raw & R1 & R2 & R3 & R4 && dev & test&& dev & test \\
    \midrule
    1    & 0    & 0    & 0    & 0     && 19.8 & 20.9 && \textbf{\textbf{6.1}} & 7.0\\
    0.8 & 0.2 & 0    & 0    & 0     && 20.1 & \textbf{\textbf{20.7}} && 6.2 & \textbf{\textbf{6.5}} \\
    0.8 & 0    & 0.2 & 0    & 0     && 20.2 & 20.9 && 6.5 & 6.7 \\
    0.8 & 0    & 0    & 0.0 & 0     && 20.1 & 21.0 && \textbf{\textbf{6.1}} & 7.0 \\
    0.8 & 0    & 0    & 0    & 0.2  && 19.8 & 20.9 && \textbf{\textbf{6.1}} & 6.6 \\
    0.8 & 0.05 & 0.05 & 0.05 & 0.05  && \textbf{\textbf{19.4}} & \textbf{\textbf{20.7}} && \textbf{\textbf{6.1}} & 6.6 \\
    0.7 & 0.15 & 0    & 0    & 0.15  && 19.8 & 20.8 && \textbf{\textbf{6.1}} & 6.6 \\
    0.4 & 0.2 & 0.2 & 0.1 & 0.1  && 21.2 & 21.8 && 7.3 & 7.9 \\
    \bottomrule
  \end{tabular}
\end{table}

In order to further study the influence of the impact of different syntax rules on the performance of ASR model, we investigate different data combination ratios for original data and augmented data.
Table \ref{tab:tab7} shows the performance of the Transformer-based ASR with different training data combination ratios under the four syntax rules.
It can be further confirmed that different syntax rules for semantic data augmentation
may have different impact, but a relatively balanced ratio setup often results in better performance.

\section{Conclusions}
\label{sec:conc}
In this work, we proposed a semantic data augmentation algorithm for end-to-end Mandarin speech recognition. 
This algorithm uses part-of-speech tagging tool CoreNLP to segment the transcriptions, and
then applies transposition strategies, such as placing the object in front of the subject or swapping the subject and the object, on the segmented sentences to give more diversity on the linguistic level.
Finally, new acoustic features with more variations 
can be reassembled based on framewise forced-alignment produced by a pre-trained GMM-HMM ASR system.
The method can produce consistent performance gain for different end-to-end ASR systems, i.e., Transformer and Conformer based systems in this paper.
In the future, we will investigate more complex syntax rules and even grammar rules for semantic data augmentation, and apply the proposed method to other languages, such as English.

\bibliographystyle{IEEEtran}

\bibliography{mybib}

\clearpage\end{CJK*}
\end{document}